\documentclass[pre,aps,twocolumn]{revtex4-1}
\pdfoutput=1
\usepackage{amsmath,natbib,amsfonts}
\usepackage{amssymb,color}
\usepackage{graphicx}
\usepackage{subcaption}
\newcommand{\beqa}{\begin{eqnarray}}
\newcommand{\eeqa}{\end{eqnarray}}
\newcommand{\lan}{\langle}
\newcommand{\ran}{\rangle}

\begin{document}

\title{Stochastic Heat Engine Powered By Active Dissipation} 
\author{Arnab Saha}
\affiliation{Department of Physics, Savitribai Phule Pune University, Ganeshkhind, Pune  411007, India.}
\author{Rahul Marathe}
\email{maratherahul@physics.iitd.ac.in}
\affiliation{Department of Physics, Indian Institute of Technology, Delhi, Hauz Khas 110016, New Delhi, India.}
\author{P. S. Pal}
\affiliation{Institute of Physics, Sachivalaya Marg, Bhubaneshwar 751005, Odhisha, India, \\ 
Homi Bhabha National Institute, Training School Complex, Anushakti Nagar, Mumbai 400085, India. }
\author{A. M. Jayannavar} 
\affiliation{Institute of Physics, Sachivalaya Marg, Bhubaneshwar 751005, Odhisha, India, \\
Homi Bhabha National Institute, Training School Complex, Anushakti Nagar, Mumbai 400085, India. }

\date{\today}
\begin{abstract}
Thermodynamics of nanoscale devices is an active area of research. Despite their noisy surrounding they often produce mechanical work (e.g. micro-heat engines), display rectified Brownian motion (e.g. molecular motors). This invokes research in terms of experimentally quantifiable thermodynamic efficiencies. Here, a Brownian particle is driven by a harmonic confinement with time-periodic contraction and expansion. The system produces work by being alternately (time-periodically) connected to baths with different dissipations. We analyze the system theoretically using stochastic thermodynamics. Averages of thermodynamic quantities like work, heat, efficiency, entropy are found analytically for long cycle times. Simulations are also performed in various cycle-times. They show excellent agreement with analytical calculations in the long cycle time limit. Distributions of work, efficiency, and large deviation function for efficiency are studied using simulations. We believe that the experimental realization of our model is possible.
\end{abstract}
\maketitle
%---------------------------------------------------------------------------------%

%\noindent{\it Introduction.}
\section{Introduction}
The Kelvin's statement of the Second Law of Thermodynamics reads ``it is impossible to extract heat from a single heat bath which can be used to perform useful work''\cite{Callen}. Heat bath provides thermal fluctuation and dissipation via frictional drag to a system (e.g. described by a Langevin equation) in contact. For heat baths in equilibrium, they are related via Einstein's fluctuation-dissipation relation (FDR)  which prohibits extraction of work from a single heat bath. This can be challenged in various ways. One way is instead of an open loop control, one can use a close-loop control over a process by the feedback mechanism. It is widely encountered in natural as well as artificial systems \cite{Astrom,Bechhoefer, Smith, Leff, Nori, seifertEPL11}. The other way, which is relevant for the current work is, instead of using the bath in equilibrium that maintains FDR, use the bath which is out of equilibrium where the dissipative processes break FDR to extract useful thermodynamic work. 

The system we consider here is a harmonically confined Brownian particle and a four-stage cyclic, open-loop protocol by which the confinement time-periodically expands and contracts. While contracting, the heat bath is driven out of the equilibrium such that the particle faces a different frictional dissipation but same thermal fluctuation as it was facing along the stage of expansion. This time-periodic modulation of friction and protocol together causes work extraction which we are going to estimate here. Similar systems, where non-equilibrium reservoirs produces athermal fluctuations to promote $(a)$ unidirectional energy transport \cite{Sriram2005} and $(b)$ work-extraction \cite{Sood16} at small scales have been studied recently. One can also extract work using micro-heat engines (e.g.\cite{Marathe07,bechinger10,Martinez16,Edgar16,Arun14}) or using micro-refrigeration techniques {\cite{ArunPhysicaA, ArunIJMPB}} where the trapped Brownian particle is time-periodically driven between two heat baths having different temperatures. In these micro-machines the heat baths are in equilibrium and therefore temperature is a well defined quantity which is not the case when the bath is driven out of equilibrium.

In our system, the frictional drag faced by the particle immersed in a heat bath is varied time-periodically. It is motivated by the usage of active systems as a non-equilibrium heat bath for micro-heat engines {\cite{Sood16}}. In experiments, the activity of bacterial heat bath can be modulated over a cycle of a micro-heat engine (attached to the bath) externally, in a time-periodic manner (for example, an assembly of phototactic bacteria with external light source, the intensity of which is oscillatory in time) without considerable alteration of thermal fluctuations. In such systems, with appropriate combinations of activity of individual entities and passive interactions (which may include both short-ranged steric and long-ranged hydrodynamic interactions) among them, large scale crystalline and other patterns may emerge {\cite{hartmut2013, Alexander2015,palacci2013}}. It has been shown experimentally {\cite{Alexander2015}} as well as theoretically {\cite{Rajesh2016}}  that in such active, non-equilibrium systems, instead of thermal energy, activity together with hydrodynamic interactions can play major role behind the emergence of large scale patterns. When a passive Brownian particle is immersed in such an active, non-equilibrium heat bath, the large scale patterns within the bath can provide a friction to the particle which is in general different from a passive heat bath at equilibrium. Therefore the dissipation from the active heat bath can be  {\it de-coupled} or {\it independent} from its thermal fluctuations. Similar effect is observed recently even when an active tracer particle is immersed in a passive bath \cite{Cates18}.  The friction from active heat bath can be large enough to {\it{suppress}} the thermal fluctuations of the Brownian particle (immersed into it) more, as compared to the friction provided by the heat bath at equilibrium. More over, the friction can be controlled externally via modulating activity of the bath. This essentially motivates us to explore the stochastic thermodynamics of micro-machines driven by the protocol described here.

We analyze our model using stochastic thermodynamics \cite{Sekimoto97,Seifert12}. We compute distributions and averages of various thermodynamic quantities - e.g. work, heat, entropy production, efficiency etc. over a large number of stochastic trajectories both in small as well as long cycle-time limit. In the following section we will explain our model. In Results section we will explain the analytical and numerical methods to solve the model equations as well as the procedure to calculate stochastic thermodynamic quantities. Then we analyze our results with physical interpretation and finally we conclude with the discussion of our results and point out possible experimental realization. 

\section{Model}
We consider a Brownian particle confined in a Harmonic trap. The trap strength is time-periodic and used as a protocol to drive the particle. The protocol used here is similar to the one used in \cite{Arun14} but with an important difference. The equation of motion of the particle, when in contact with the heat bath equilibrated at temperature $T$, is given by the under-damped Langevin equation:
\beqa
m\ddot{x} = -\gamma \dot{x} -k(t)x+\sqrt{D} \xi(t).
\label{Lang1}
\eeqa
Here, $m$ is the mass of the particle, $\gamma$ the friction coefficient, $T$ the temperature of the bath, $k(t)$ is time-dependent trap strength and $D=2\gamma k_BT$. The noise $\xi(t)$ comes from the heat bath and modeled as Gaussian white noise satisfying $\lan\xi(t)\ran=0$ and $\lan\xi(t)\xi(t')\ran=\delta(t-t')$. In all the calculations we keep mass $m$ and the Boltzmann constant $k_B$ as unity.  All the figures are also plotted in the units of $k_BT$. The strength of the confinement $k(t)$ is varied with time in a cycle of duration $\tau$. This protocol undergoes following steps. In the first step $k(t)$ is decreased linearly from the initial value $k$ to $k/2$ as:
\beqa
k(t)=k\left(1-\frac{t}{\tau}\ \right)=k_1(t). ~~~~~~~0<t<\tau/2 \nonumber
\eeqa
This step is the expansion step with bath temperature $T$. After this the trap strength is decreased further to $k/4$ instantaneously. In the third step $k(t)$ is increased linearly from $k/4$ to $k/2$ as:
\beqa
k(t)=k\frac{t}{2\tau}\ =k_2(t), ~~~~~~~\tau/2<t<\tau \nonumber
\eeqa
In this compression step, the heat bath is out of equilibrium and we assume that the non-equilibrium processes of the bath induces more dissipation and negligible fluctuations to the system.  Therefore, unlike Eq. \ref{Lang1}, effective friction coefficient increases from $\gamma$ to $\gamma_{eff}=\gamma + \gamma_{a}$ (with $\gamma_{a} > 0$) and the equation of motion of the particle becomes:  
\beqa
m\ddot{x} = -(\gamma_{a}+\gamma) \dot{x} -k(t)x+\sqrt{D} \xi(t).
\label{Lang2}
\eeqa
Here $\frac{D}{2\gamma}=k_BT$ represents energy scale related only to the thermal fluctuations in the bath. Due to $\gamma_a$, FDR is broken and the bath goes far from equilibrium. This is the crucial step which allows the particle to cool down by the excess amount of friction and thereby to extract heat from the bath that can be used to perform useful work. We note here that defining the effective temperature as $T_{eff}=\frac{T\gamma}{\gamma+\gamma_a}$, FDR can be restored. We will come back to this point later, while discussing average efficiency of the system.
 
\begin{figure}[!thbp]
\centering
\includegraphics[width=0.80\columnwidth]{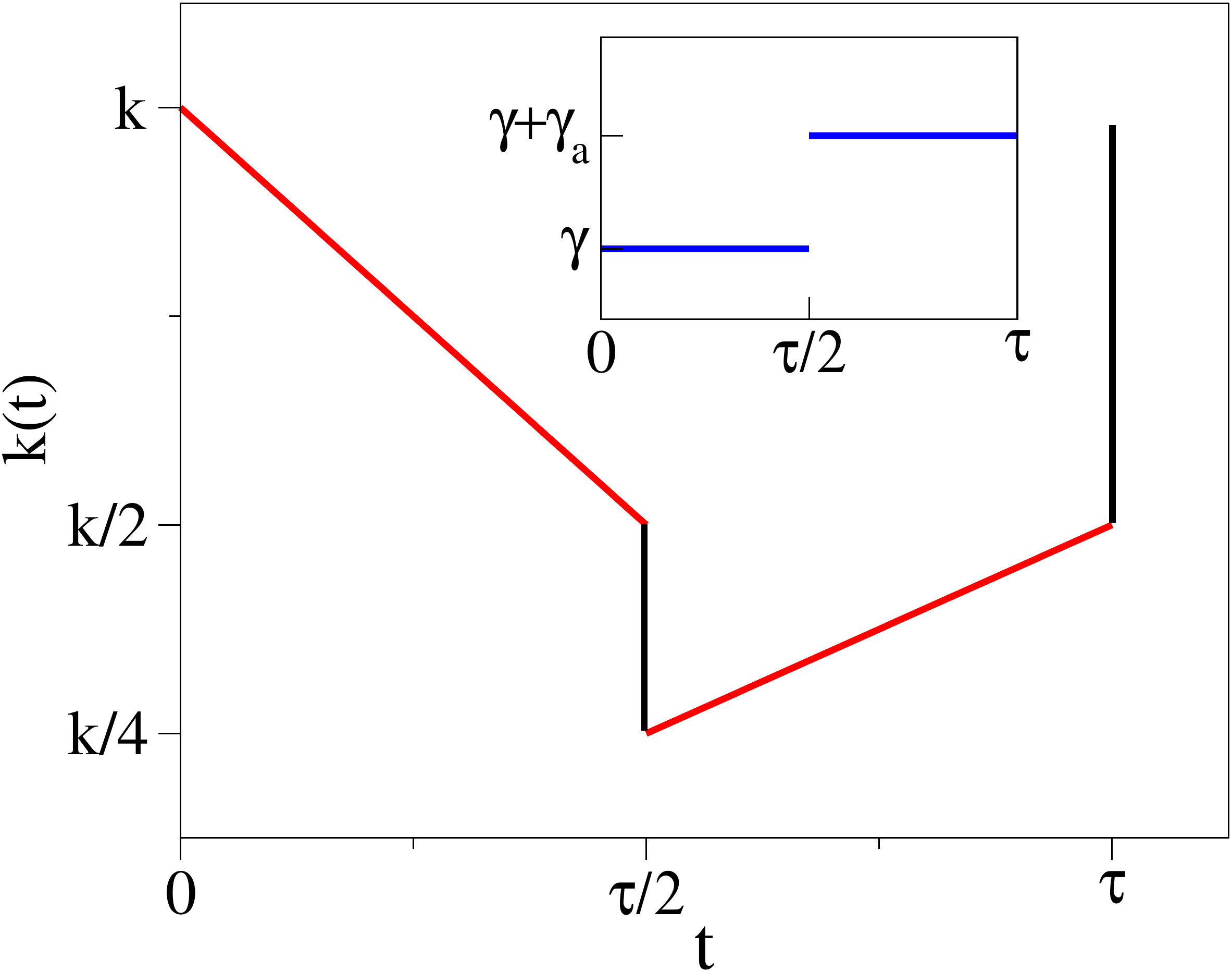}
\caption{The schematic diagram  of the protocol: Red lines imply the isothermal, gradual change of the protocol i.e. the strength of the trap. From $t=0$ to $t=\tau/2$ it decreases with friction coefficient $\gamma$ and from $t=\tau/2$ to $t=\tau$ it increases with effective friction coefficient $\gamma+\gamma_{a}$ to drive the bath out of equilibrium. The black lines of the protocol imply sudden jumps. Inset shows the variation of the friction coefficient.}
\label{prot}
%\vspace{-3mm}
\end{figure}

In the last step, the trap strength is finally increased to its initial value $k$ from $k/2$ instantaneously. In the beginning of this step, the friction coefficient is restored to $\gamma$ so that the system is ready for the next expansion step. This cycle is then be repeated. The protocol is depicted in Fig. \ref{prot}.

In the introduction it is mentioned that the protocol described above is motivated by extracting thermodynamic work from active systems used as active bath for micro heat engines. It should be noted here that thermodynamics of active systems, which is inherently out of equilibrium, cannot be described with effective temperature {\it{unconditionally}}. For example, in case of dilute suspension of run and tumble bacterium, whenever the external force cannot perturb their swim
speed considerably and cannot provide sharp gradients within its velocity field over the scale of their run-length, then only the thermodynamics of such suspension can be described by effective temperature \cite{Cates09}. In general, wherever it is possible to have active systems with Gaussian fluctuations leading towards Boltzmann-like distribution with an effective temperature, different from the bath temperature, in such cases the current research is applicable.

It should also be noted here that in case of micro heat engines, there are multiple ways to implement adiabaticity along a thermodynamic process. For example, one can design a special cyclic protocol that keeps the phase space volume of the system constant in quasistatic limit (and therefore no average heat dissipation) along a particular step, named as {\it{microadiabatic}} step \cite{Martinez2015}. Another way to implement adiabatic condition, is implementing cyclic protocols that connects two different isotherms of the cycle by instantaneous jumps of the protocol \cite{Seifert2008, Arun14}. In overdamped limit, the fact that along the instantaneous jumps the system does not find enough time to release or absorb heat from the bath, leads to iso-entropicity and consequently adiabaticity. Though in case of under-damped dynamics as we follow here, due to the kinetic energy part, the sudden change in internal energy of the particle occurred during the quenches, will contribute to heat which is released or absorbed when the system relaxes along the isotherms immediately after the instantaneous quenches \cite{Lutz18}. These contributions will be considered in our calculation accordingly.

We are interested in quantifying stochastic thermodynamical quantities and there averages, like total work done along a trajectory : $W=W_1+W_2+W_3+W_4$, where $W_i$'s are work performed in {\it i-th} step of the protocol, and heats exchanged between the system and the bath, denoted by $Q_i$'s during these steps. Internal energy of the Langevin system is given by $U(x,\dot{x})=\frac{1}{2}\ m\dot{x}^2 +\frac{1}{2}\ k(t) x^2$. Using Stochastic Thermodynamics \cite{Sekimoto97,Seifert12}, we can also find out expressions for work done and heat exchanged in all four steps of the protocol. In the expansion process work done on the system and heat absorbed by the system are given by $W_1=\int_{0}^{\tau/2}\frac{1}{2}\ \dot{k}_1(t) x^2(t) ~dt$ and $Q_1=\int_{0}^{\tau/2}(-\gamma \dot{x} +\sqrt{D}\xi(t)) \dot{x}(t)~dt$ respectively. In the instantaneous expansion step no heat exchange takes place $(Q_2=0)$ and work done on the system is nothing but change in the internal energy, i.e. $W_2 = \frac{1}{2} (k_2(\tau/2)-k_1(\tau/2))x^2(\tau/2)$. In the third step again work and heat definitions remain as in the first step but with modified friction term and thus $W_3=\int_{\tau/2}^{\tau}\frac{1}{2}\ \dot{k}_2(t) x^2(t) ~dt$ and $Q_3=\int_{\tau/2}^{\tau}(-(\gamma_{a}+\gamma) \dot{x} +\sqrt{D}\xi(t)) \dot{x}~dt$. Fourth step, being instantaneous again, gives no heat exchange $(Q_4=0)$ but work done on the system is given by the change in the internal energy as, $W_4  = \frac{1}{2} (k_1(0)-k_2(\tau))x^2(\tau)$. According to our sign convention the work done on the system and heat absorbed by the system are positive. First we calculate work and change of internal energy along each trajectory of the particle, following the definitions above. Then we apply the first law of stochastic thermodynamics, where the difference of internal energy $\Delta U_i(x,\dot x)=W_i+Q_i$ for the $i^{th}$ step, to calculate the heat exchange, instead the integral expressions of $Q_i$'s, mentioned before. One can be sure about the validity of first law by considering the infinitesimal heat exchange between the particle and the bath as $dq=(-\gamma\dot x+\sqrt{D}\xi)\dot x dt=(m\ddot x+k(t)x)\dot x dt$. Here we have used the equation of motion of the particle and considered the system is in isothermal expansion mode. When the system is in compression mode, the protocol and friction will be changed accordingly but the argument here to validate first law is generic enough to apply in both cases. It is straight forward to write the infinitesimal heat as $dq=\frac{d}{dt}(\frac{1}{2}m\dot x^2+\frac{1}{2}k(t)x^2)dt-\frac{1}{2}\dot kx^2dt=du-dw$ where we have identified the first term with infinitesimal change in internal energy $du$ and second term as infinitesimal thermodynamic work $dw$ done within the infinitesimal time $dt$. For instantaneous jumps $dq=0$ and therefore $du=dw$. This is the first law in stochastic thermodynamics \cite{Sekimoto97,Seifert12}, in the context of a single trajectory of the particle. This can be integrated over time and averaged over realizations to evaluate average heat dissipation in the expansion or compression step of the protocol here.

Using the definitions of heat and work discussed above, we define stochastic efficiency over a single trajectory of the Brownian particle as, $\eta=-W/Q_1$. 
Two different averages of $\eta$ over the cycles can be calculated as:

\beqa
\bar\eta = \frac{\lan -W\ran}{\lan Q_1\ran }\ ,~~~ \lan\eta\ran = \left\lan\frac{-W}{Q_1}\ \right\ran ,
\label{etadef}
\eeqa
where the angular brackets imply steady state average over all possible realizations. Later we will see that the distribution of $\eta$ has a power law tail with power close to $-2$. Therefore $\lan\eta\ran$ is not a well defined quantity.

Next we present results obtained analytically in long cycle time limit and also by simulating the system in both long as well as short cycle times. We calculate thermodynamic quantities from simulation for very long cycle time to compare the results we obtain analytically. This will indicate us that for a given set of $\gamma$ and $\gamma_a$, for which $\tau$ the system departs from its long cycle time behavior. We will calculate distributions of thermodynamic work and stochastic efficiency in both long as well as short cycle time regime from simulation and then we numerically calculate large deviation function for stochastic efficiency distribution. 

In simulations we integrate the Langevin equations Eq. (\ref{Lang1}) or Eq. (\ref{Lang2}), depending on whether the trap is expanding or contracting, by a velocity Verlet algorithm with Stratonovich discretization having time step $dt\sim 10^{-3}$ and find average work and heat exchanged. These averages are taken over $10^5$ cycles of $k(t)$, after driving the system in the steady state.

\begin{figure*}[!thbp]
\begin{subfigure}{0.4\linewidth}
\centering
\includegraphics[width=7.5cm,height=5.5cm,angle=0.0]{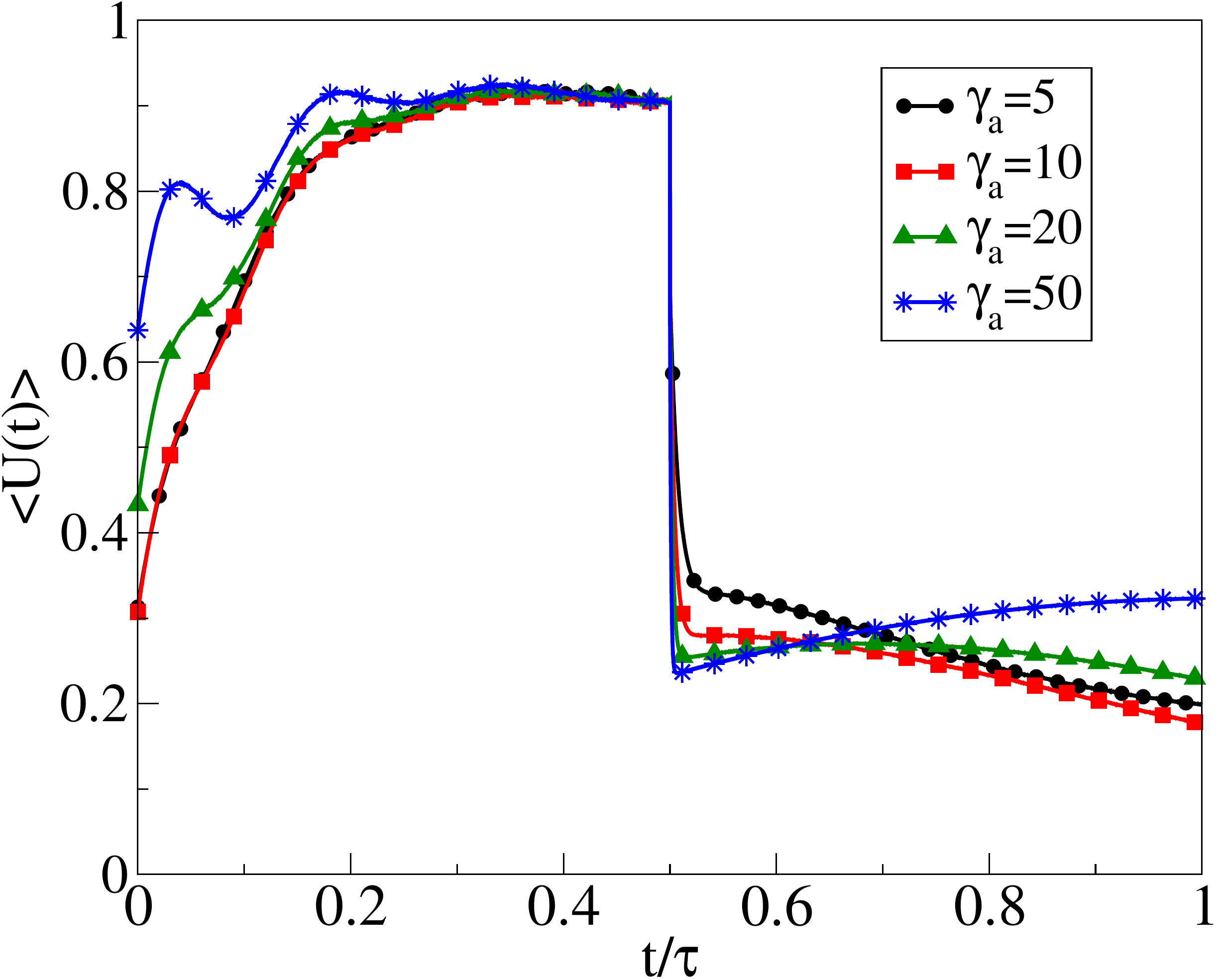}
\caption{}
\end{subfigure}
\hspace{1cm}
\begin{subfigure}{0.4\linewidth}
\centering
\includegraphics[width=7.5cm,height=5.5cm,angle=0.0]{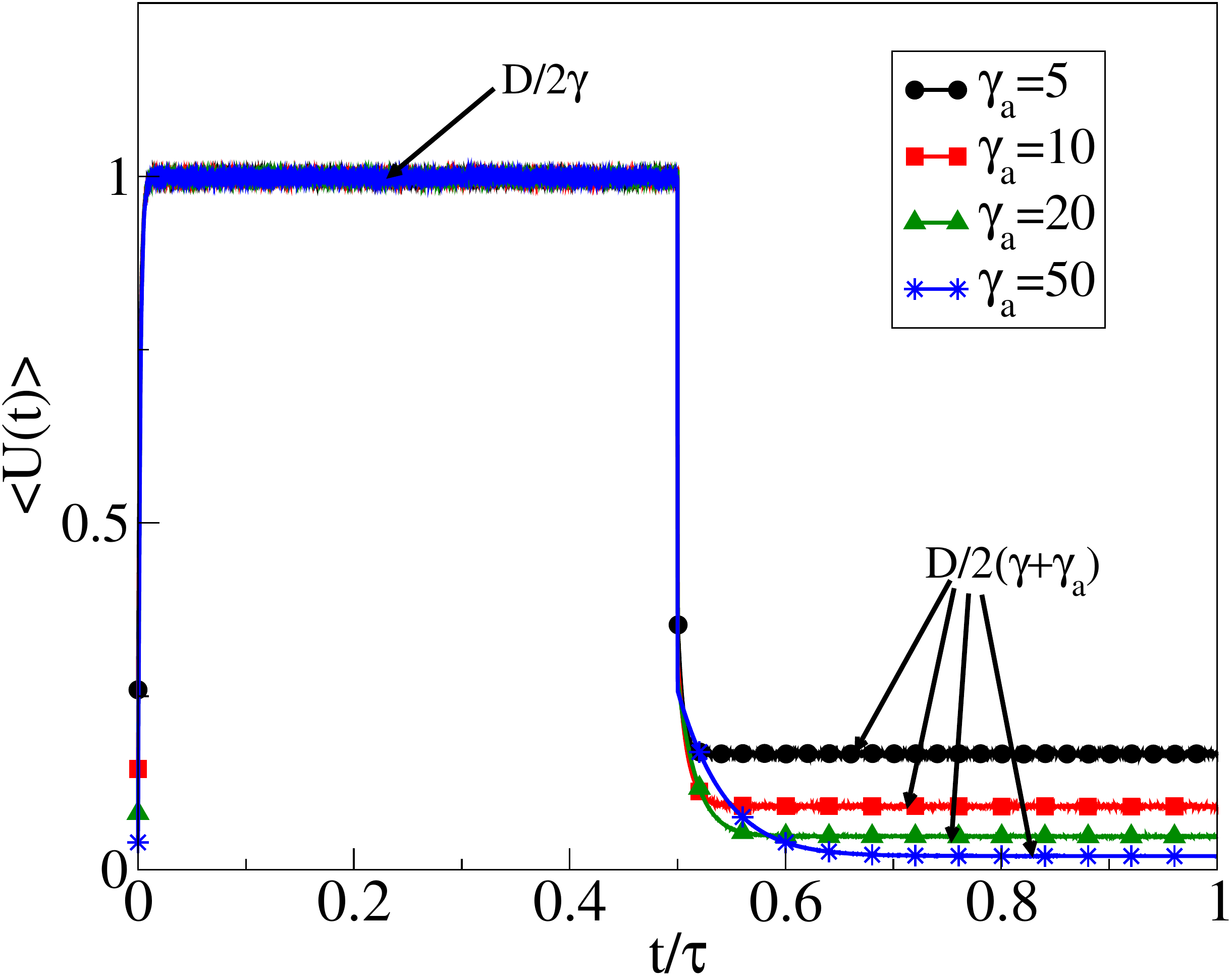}
\caption{}
\end{subfigure}
\caption{Evolution of average internal energy along the driving cycle as a function of $t/\tau$,
for different $\gamma_a$'s. Left panel $\tau=10$, right panel $\tau=500$. Note that in long cycle time limit, i.e.  
$\tau=500$, energy saturates to corresponding temperatures in respective half-cycles of the protocol. 
Parameters used are $T=1$, $\gamma=1$, $k_0=5$.}
\label{endiffgama}
\end{figure*}

\section{Results}

\subsection{Average Thermodynamic Quantities }

In this section we calculate the average thermodynamic quantities (e.g. work, heat, efficiency etc.) where $\tau$ is much longer time scale in comparison to any other time scale present in the problem. The average is the ensemble average. We begin by writing the dynamics of velocity and position fluctuations derived from the equation of motion of the particle along isothermal compression (i.e. Eq.[\ref{Lang1}]) as \cite{Lutz18}:

\beqa
&&\frac{d\sigma_v}{dt} = -2\gamma\sigma_v-k_1(t)\frac{d\sigma_x}{dt}+D 
\label{SigmaDynamics_v}
\eeqa
\beqa
&&\frac{d^2\sigma_x}{dt^2}=2\sigma_v-\gamma\frac{d\sigma_x}{dt}-2k_1(t)\sigma_x
\label{SigmaDynamics_x}
\eeqa
were $\sigma_v=\langle v^2\rangle-\langle v\rangle^2$ and $\sigma_x=\langle x^2\rangle-\langle x\rangle^2$. As we are driving the particle simply by changing the spring constant of the harmonic trap with time, keeping the mean position and velocity of the particle constant, without losing generality we fix $\langle x\rangle=0, \langle v\rangle=0$. In case of isothermal compression the mathematical form of above equations will be same except the fact that $\gamma$ will be replaced by $\gamma+\gamma_a$ and $k_1(t)$ will be replaced by $k_2(t)$. In Eq.[\ref{SigmaDynamics_v}], we change the variable $t \rightarrow t/\tau$ and then take $\tau\rightarrow \infty$ limit to obtain the velocity fluctuation in isothermal expansion and compression as:
\beqa
\sigma_v=\frac{D}{2\gamma} \phantom{xxx} ({expansion}), ~\text{and,} \nonumber \\ 
\sigma_v=\frac{D}{2(\gamma+\gamma_a)} \phantom{xxx} ({compression}).
\label{sigv}
\eeqa
Position fluctuation of the particle along isothermal expansion and compression in $\tau\rightarrow \infty$ limit can similarly be extracted from Eq.[{\ref{SigmaDynamics_x}}] as:
\beqa
\sigma_x&=&\frac{D}{2\gamma k_1(t)}  \phantom{xxx} ({expansion}), ~\text{and,} \nonumber \\ 
\sigma_x&=&\frac{D}{2(\gamma+\gamma_a)k_2(t)} \phantom{xxx} ({compression}).
\label{sigx}
\eeqa  

Here, while calculating the thermodynamic quantities such as work and heat in $\tau\rightarrow\infty$ limit, we will use Eq. [\ref{sigv}, \ref{sigx}]. Note that here though the protocol jumps instantaneously, $\sigma_x$ and $\sigma_v$ remain continuous functions of time through out the cycle. Therefore at $t=\tau$ and at $t=\tau/2$ also, they are continuous, i.e.  $\sigma_{(x,v)}|_{\tau^-}=\sigma_{(x,v)}|_{\tau^+}=\sigma_{(x,v)}|_{\tau}$ and $\sigma_{(x,v)}|_{(\tau/2)^-}=\sigma_{(x,v)}|_{(\tau/2)^+}=\sigma_{(x,v)}|_{\tau/2}$. Thus, while calculating the difference of internal energies before and after the quenches, we can use $\sigma_{(x,v)}|_{\tau/2}$ for instantaneous expansion and $\sigma_{(x,v)}|_{\tau}$ for instantaneous compression. In numerics we will estimate thermodynamic quantities (e.g. heat, work etc.) both in short and long time regimes. We will see that, the analytical results obtained in $\tau\rightarrow\infty$ limit together with the continuity of fluctuations, can indeed be recovered in numerics for long cycle times i.e. for cycle times much larger than the relaxation time of the system. To estimate the relaxation time approximately we consider the dynamics of $\sigma_x$ after eliminating $\sigma_v$ terms, in Eq. $(\ref{SigmaDynamics_x})$, upto $\mathcal {O} (\tau^{-1})$ which gives the relaxation time $\tau_r\simeq (2k+\gamma^2)/(2\gamma k+\dot{k}(t)) $, where $\dot{k}(t)=\frac{dk}{dt}$ and $k$ is the value of spring constant at time $t=0$. Similar estimation can be made for the second isotherm with dissipation $\gamma+\gamma_a$. For long cycle time limit we need $\tau >>\tau_r$.

\begin{figure*}[!htbp]
\begin{subfigure}{0.3\linewidth}
\centering
\includegraphics[width=5cm,height=4cm,angle=0.0]{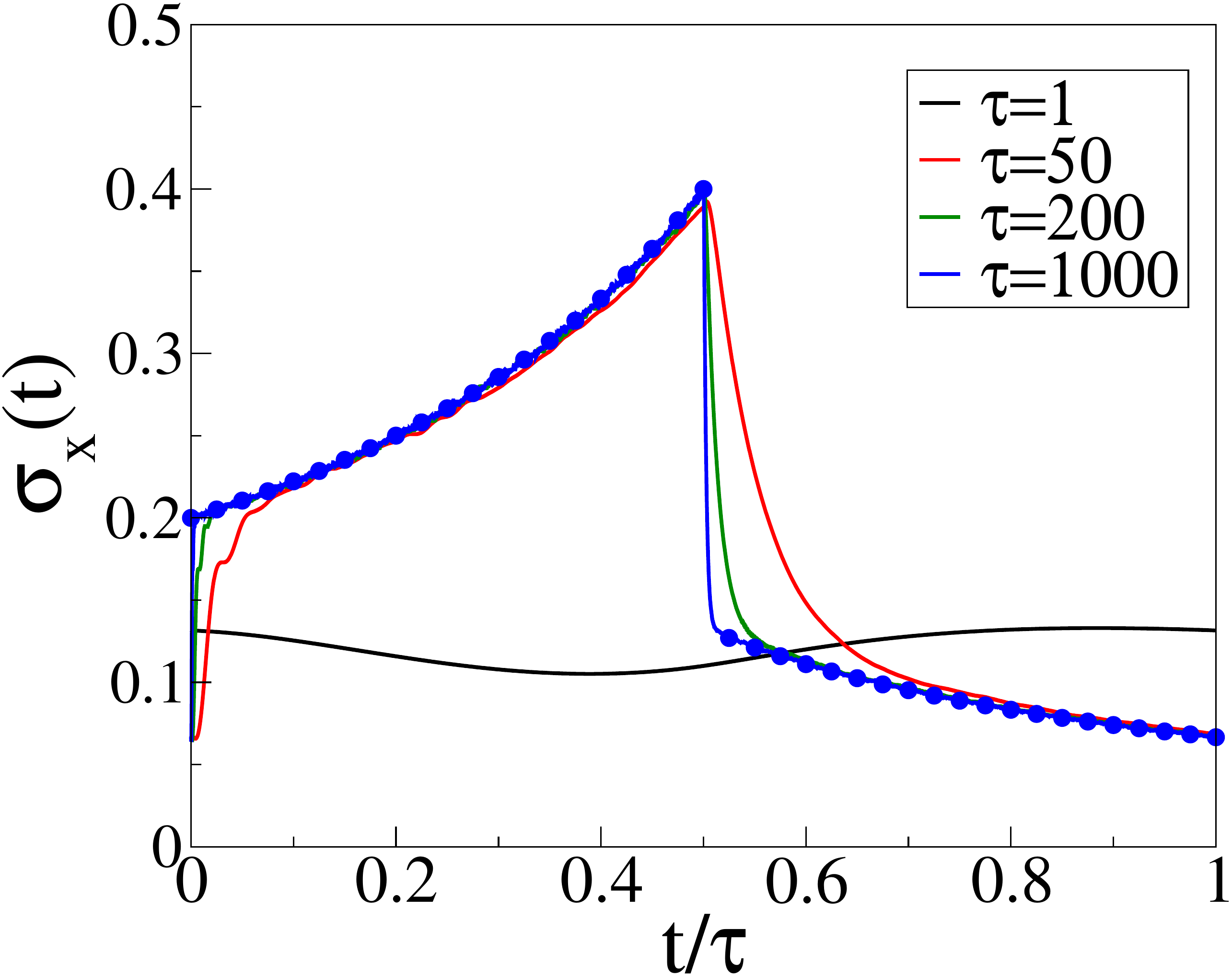}
\caption{}
\end{subfigure}
\begin{subfigure}{0.3\linewidth}
\centering
\includegraphics[width=5cm,height=4cm,angle=0.0]{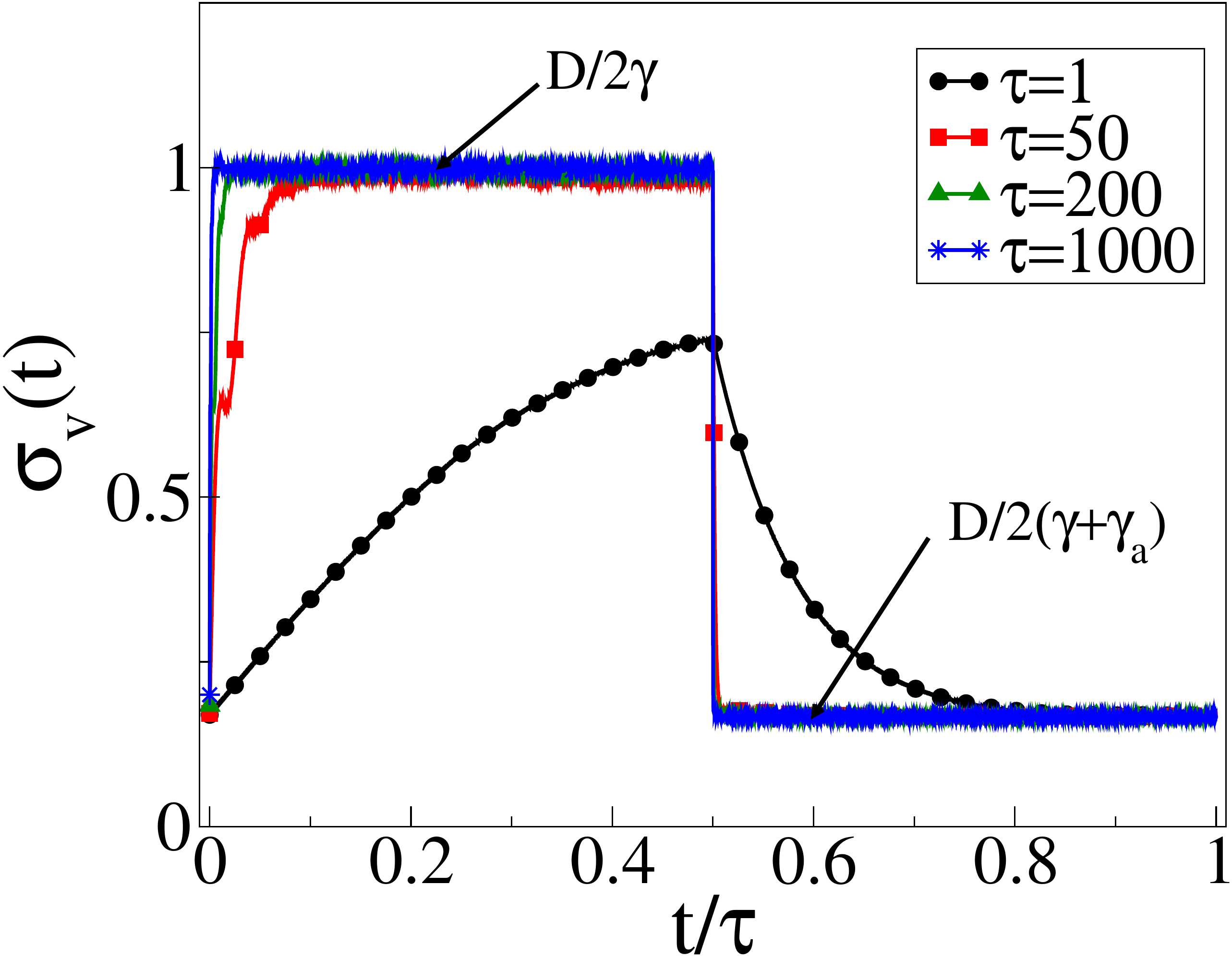}
\caption{}
\end{subfigure}
\begin{subfigure}{0.3\linewidth}
\centering
\includegraphics[width=5cm,height=4cm,angle=0.0]{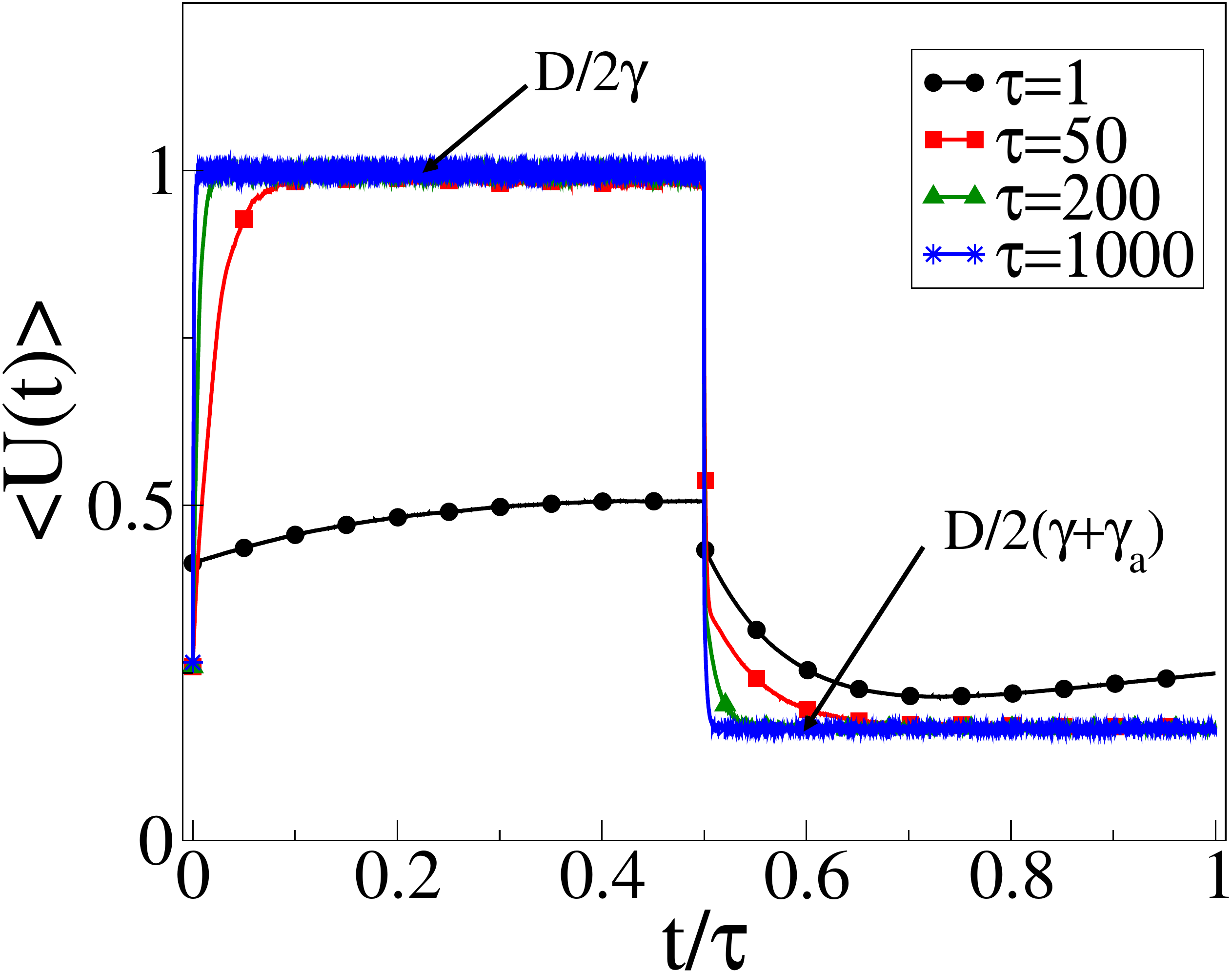}
\caption{}
\end{subfigure}
\caption{Evolution of $(a)$ $\sigma_x$, $(b)$ $\sigma_v$ and $(c)$ average internal energy, along the 
driving cycle as a function of $t/\tau$, for different cycle times $\tau$. Note that 
for $\tau=1000$, $\sigma_x(t)$ (panel $(a)$) saturates to analytical results in Eq. $(\ref{sigx})$ 
(blue filled circles) and $\sigma_v$ (Eq. $(\ref{sigv})$) and also the total energy saturates to corresponding temperatures in respective half-cycles 
of the protocol. We have used $T=1$, $\gamma=1$, $\gamma_a=5$, $k_0=5$.}
\label{sigxenw}
\end{figure*}

\begin{figure*}[!htbp]
\begin{subfigure}{0.4\linewidth}
\centering
\includegraphics[width=7cm,height=5cm,angle=0.0]{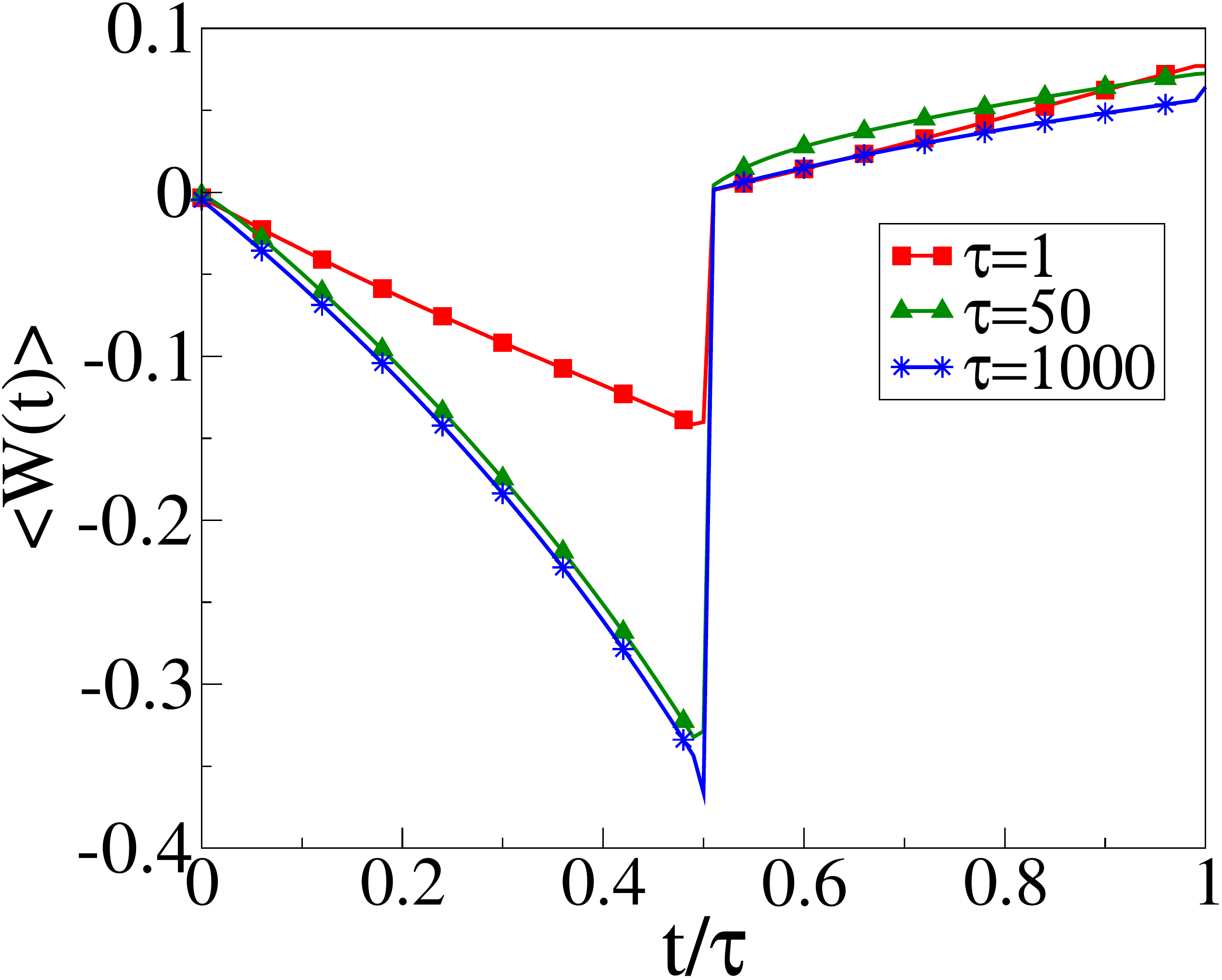}
\caption{}
\end{subfigure}
\begin{subfigure}{0.4\linewidth}
\centering
\includegraphics[width=7cm,height=5cm,angle=0.0]{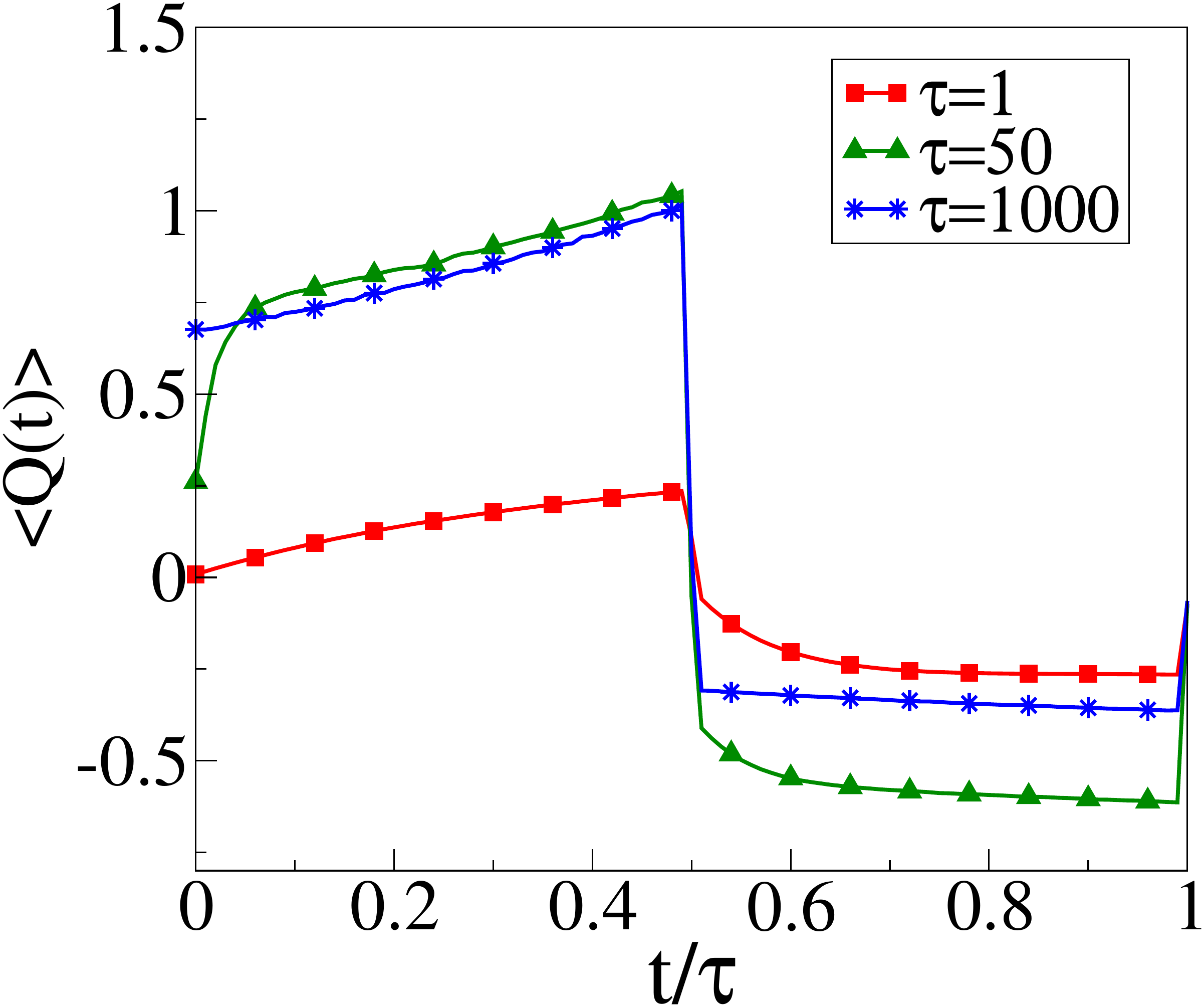}
\caption{}
\end{subfigure}
\caption{Average accumulated $(a)$ work and $(b)$ heat, upto time $t$, for different time cycles of the protocol. We have used $T=1$, $\gamma=1$, $\gamma_a=5$, $k_0=5$.}
\label{accuwq}
\end{figure*}

To inspect how long cycle time limit can be achieved in numerics, let us look at the evolution of $\sigma_x(t)$, $\sigma_v(t)$, average internal energy $(\langle U\rangle=\frac{1}{2}m\sigma_v(t)+\frac{1}{2}k(t)\sigma_x(t))$ over a period of the protocol. We plot evolution of average internal energy along $t/\tau$ for short cycle time regime i.e. for $\tau=10$, (Fig. $\ref{endiffgama}(a)$) and in long cycle time regime, i.e. for $\tau=500$, (Fig. $\ref{endiffgama}(b)$), with different $\gamma_a$. One can clearly see how the analytical results in $\tau\rightarrow\infty$ limit is recovered in numerics for long cycle time. Similarly, in Fig. $\ref{sigxenw}(a,b)$, 
$\sigma_x(t)$ and $\sigma_v(t)$ are plotted with $t/\tau$ that matches with the analytical expressions in Eq. \ref{sigv} and \ref{sigx} for long cycle time. Fig. $\ref{sigxenw}(c)$ contains the plot of average total energy with $t/\tau$ for varying $\tau$. For long cycle time they match with analytical predictions, validating the analytical results. We also plot average accumulated work and heat as a function of $t/\tau$ in Figs. $\ref{accuwq}$ $(a)$ and $(b)$ respectively. 
Accumulated work is calculated by dividing each half cycle of the protocol in finite number of intervals $t_i$'s of equal lengths, then, $\langle W_1(t_i) \rangle =\frac{1}{2}\ \int_0^{t_i} \dot{k}_1(t) \sigma_x(t)~dt $ for $0 < t_i \leq \tau/2$ and $\langle W_2(t_i) \rangle =\frac{1}{2}\ \int_{\tau/2}^{t_i} \dot{k}_2(t) \sigma_x(t)~dt $ for $\tau/2 < t_i \leq \tau$. Accumulated heat is calculated using the first law that is $\langle Q_1(t_i) \rangle = \langle U(t_i)-U(0)\rangle -\langle W_1(t_i) \rangle $ and $\langle Q_2(t_i) \rangle = \langle U(t_i)-U(\tau/2)\rangle-\langle W_2(t_i) \rangle $. In Fig. \ref{accuwq} intervals $t_i$'s are shown by filled solid symbols.

Using the expressions of $\sigma_x$ and $\sigma_v$, the average thermodynamic work along a cycle in $\tau\rightarrow\infty$ limit is given by $\langle W\rangle=\langle W_1\rangle + \langle W_2\rangle +\langle W_3\rangle +\langle W_4\rangle$ where the first and third terms represents average contributions are from isothermal expansion and compression whereas average instantaneous expansion and compression are represented by second and fourth terms respectively. For the jumps, we note that the thermodynamic work is simply the change in internal energy of the particle. Therefore, using the stochastic definition of thermodynamic work, mentioned previously, the expression for total work becomes:
\beqa
\nonumber \langle W\rangle&=&\frac{1}{2}\int_0^{\tau/2}\dot  k\sigma_xdt + \frac{1}{2}\left[k_2\left(\frac{\tau}{2}\right)-k_1\left(\frac{\tau}{2}\right)\right]\sigma_x\left(\frac{\tau}{2}\right)\\ \nonumber &+&\frac{1}{2}\ \int_{\tau/2}^{\tau}\dot k\sigma_xdt + \frac{1}{2}(k_1(0)-k_2(\tau))\sigma_x(\tau)\nonumber\\
\label{totw1}
\eeqa
where $\sigma_x(\tau)=\sigma_x(0)$ due to periodicity. From above, the expressions for average works done along different steps of the cycle are:
\beqa
\langle W_1\rangle&=&-\frac{k_BT}{2}\ \ln(2), \nonumber \\
\langle W_3\rangle&=&\frac{k_BT}{2}\ \left(\frac{\gamma}{\gamma+\gamma_a}\ \right) \ln(2), \nonumber \\
\langle W_2\rangle&=&-\frac{k_BT}{4}\ ,\nonumber \\
\langle W_4\rangle&=&\frac{k_BT}{2}\ \left(\frac{\gamma}{\gamma+\gamma_a}\ \right),\nonumber 
\eeqa
and therefore the total average work in this limit becomes: 
\beqa
\langle W\rangle=\frac{k_BT}{2~(\gamma+\gamma_a)} \left[\left(\frac{\gamma-\gamma_a}{2}\right) -\gamma_a \ln(2)\right]
\label{totw}
\eeqa  
Above expressions are derived with $\tau\rightarrow\infty$ limit of the cyclic process together with the continuity of $\sigma_x$ and $\sigma_v$ through out the cycle, including the jumps.

\begin{figure}[!tbhp]
\centering
\includegraphics[height=6cm,width=7.5cm,angle=0]{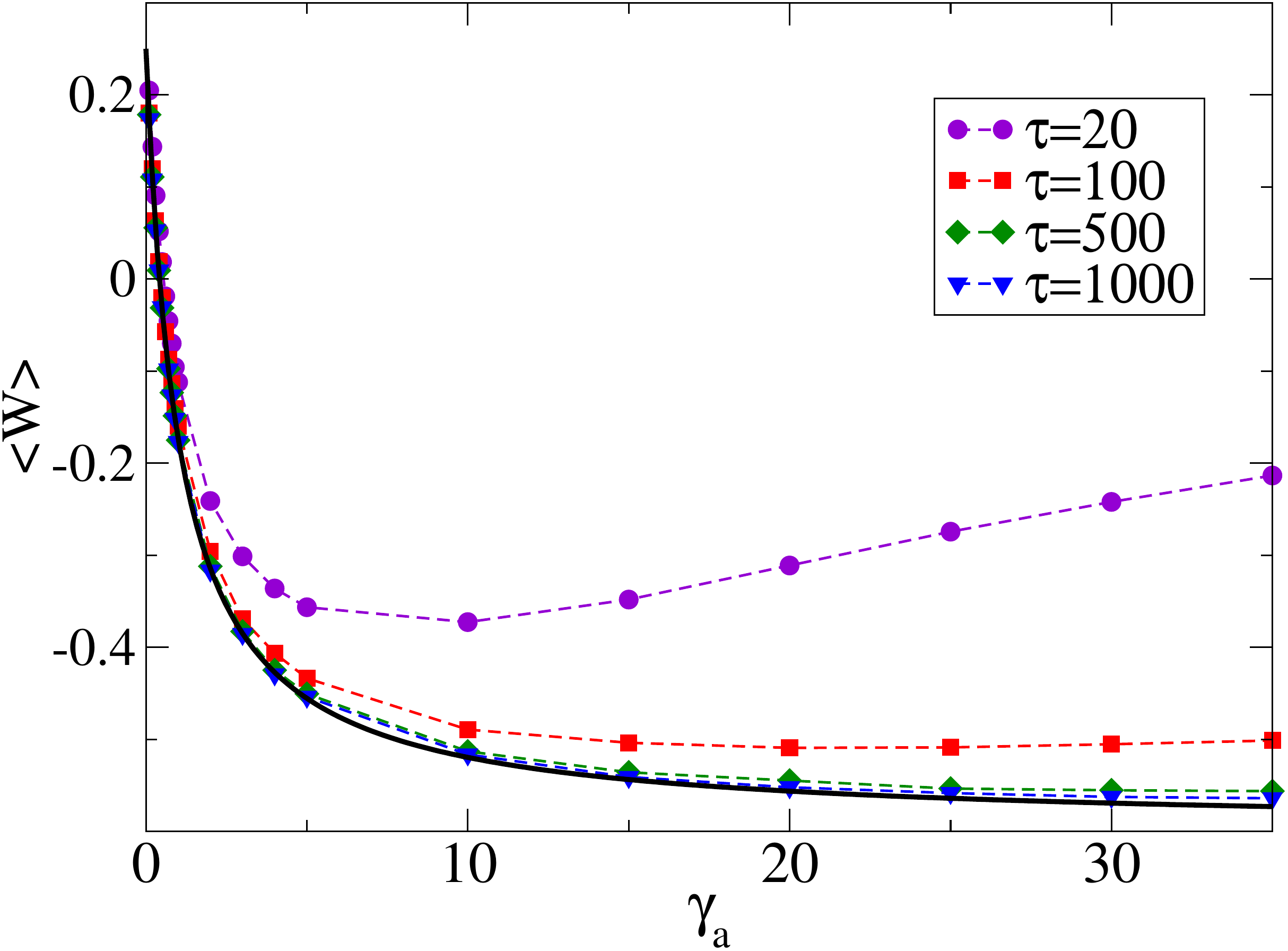}
\caption{Plot of total work $\langle W \rangle$ as a function of active friction $\gamma_a$ for different time periods.
One can see as $\tau\sim 10^3$, the results obtained analytically, is reached (solid black line). Also note that for initial values of $\gamma_a$ work is not extracted because the condition $\gamma < \gamma_a (1+2\ln(2))$  is not satisfied as discussed in the text. This plot also shows that the short cycle time behavior is very different from the long cycle time behavior (for example plot for $\tau=20$). Here we have used $\gamma=1$, $k=5$, $T=1$.}  
\label{wvsgam}
\end{figure}
\begin{figure}[!tbhp]
\centering
\includegraphics[height=6cm,width=7.5cm,angle=0]{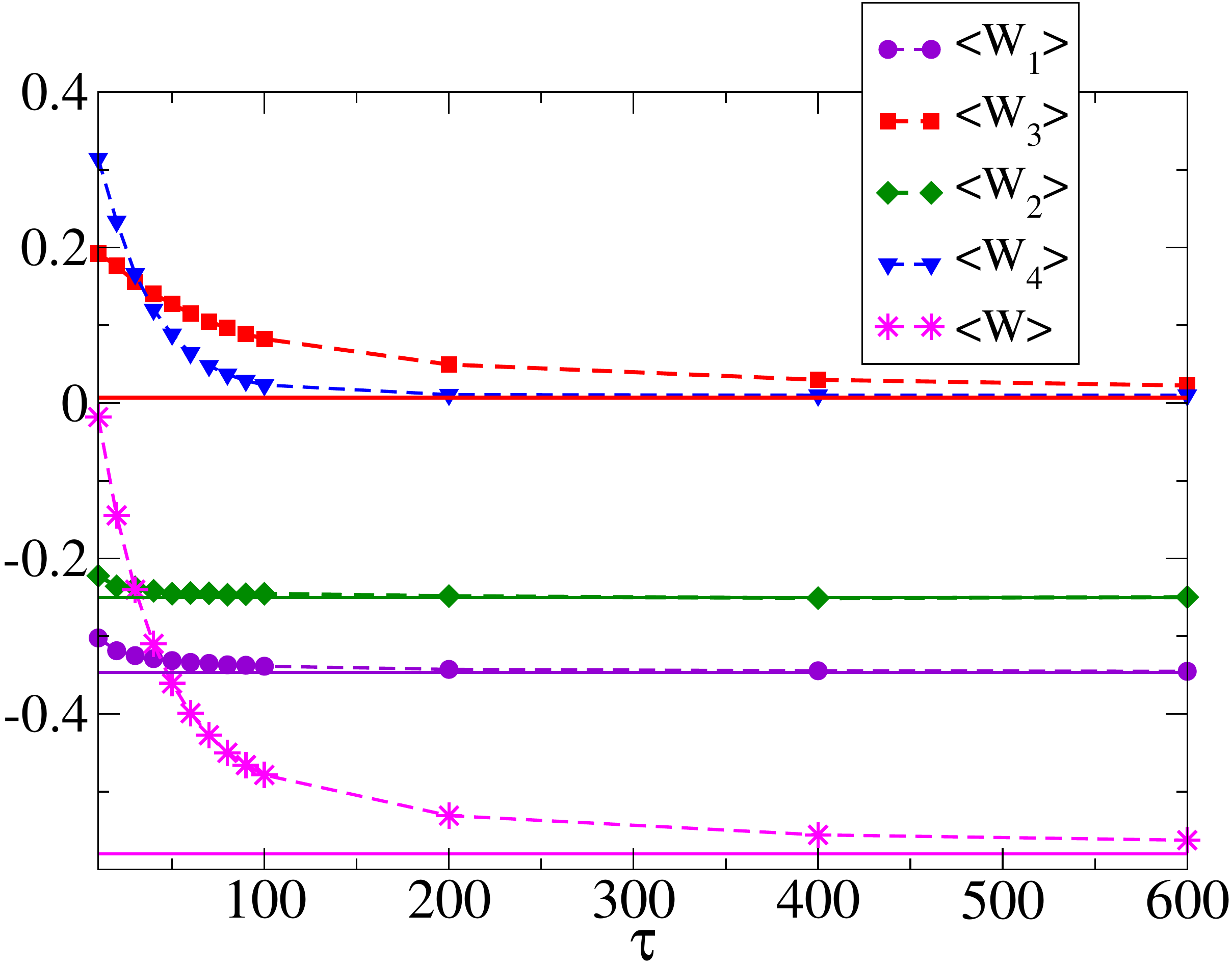}
\caption{Plot of different works in four different steps of the protocol, as a function of the time period $\tau$. As $\tau$ increases we recover analytically obtained results (solid lines), as discussed in the text. Here we have used $\gamma=1$, $\gamma_a=50$, $k=5$, $T=1$.} 
\label{wvstau}
%\vspace{-3mm}
\end{figure}

From Eq. \ref{totw}, it can be easily seen that for $\gamma < \gamma_a (1+2\ln(2))$, work can be extracted from the system and it works as an engine. In particular when $\gamma_a>>\gamma$, $\langle W\rangle\simeq -0.6k_BT$. Therefore in this limit, $0.6k_BT$ is the maximum work that can be extracted. When $\gamma_a<<\gamma$, $\langle W\rangle\simeq +\frac{k_BT}{4}\ $ and therefore work cannot be extracted from the system rather should be exerted on the system to drive it slowly. In this regime the device is not acting as an engine. Work can neither be extracted nor be performed on the system when $\gamma = \gamma_a (1+2\ln(2))$. For this combination of frictions, $\langle W\rangle=0$. In Figs. \ref{wvsgam} and \ref{wvstau}, we compare the analytically obtained long cycle time behavior with the simulation results. 

It can be noted here that usually in long cycle time limit the change in friction coefficient due to the change in viscosity of the bath alters the relaxation time scale of the system but not the mean value of a thermodynamic quantity like work. But here it does. This is because here the change in friction coefficient breaks FDR through out half of the cycle for any cycle time and therefore it becomes possible to extract work out of the system that depends on the friction coefficient explicitly. However, in the limit $\gamma_a >> \gamma $ it becomes independent of $\gamma_a$ as discussed earlier.

Now we will evaluate average heat exchanged between the system and the bath through out a cycle in long cycle time limit. As the engine is running by a cyclic protocol, the change in average internal energy of the particle is zero after completing a cycle. For half of the cycle, say for isothermal expansion, from first law of thermodynamics, average heat exchange $\langle Q_1\rangle=-\langle W_1\rangle+\langle U(\tau/2)\rangle-\langle U(0^-)\rangle$ where $\langle U(t)\rangle$ is the average internal energy of the particle at time $t$. Note that, to estimate the contribution of the sudden change in internal energy during the instantaneous jump at $t=\tau$ to $\langle Q_1\rangle$, we need to estimate $U$ from $t=0^-$ (due to periodicity, it is same as $t=\tau^-$.). Thus, using the expressions for velocity and position fluctuations derived for long cycle time limit, one can obtain:   
\beqa
\nonumber \langle Q_1\rangle&=&-\frac{1}{2}\int_0^{\tau/2}\dot  k\sigma_xdt\\ \nonumber
&+&\frac{1}{2}\sigma_v\left(\frac{\tau}{2}^-\right)+\frac{1}{2}\frac{k}{2}\sigma_x\left(\frac{\tau}{2}^-\right)-\frac{1}{2}\sigma_v\left(0^-\right)-\frac{1}{2}k\sigma_x\left(0^-\right)\\
&=& \frac{k_BT}{2}\left(\frac{2\gamma_a-\gamma}{\gamma+\gamma_a}+\ln(2)\right).
\eeqa 

\begin{figure}[!tbhp]
\centering
\includegraphics[height=6cm,width=7.5cm,angle=0]{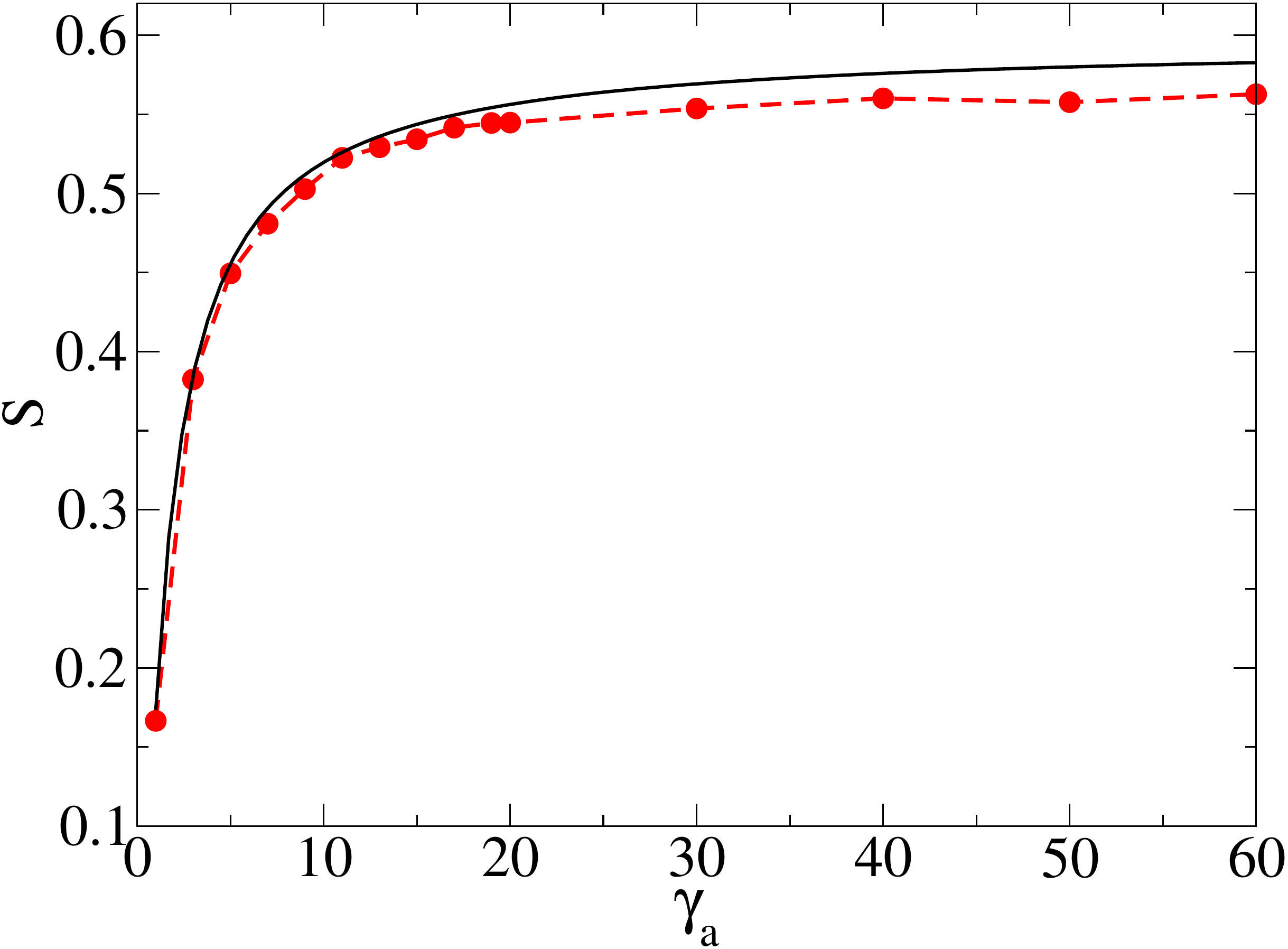}
\caption{Plot of change of entropy along a cycle with $\gamma_a$. Solid line represents Eq. $(\ref{entropy})$. 
Other parameters are $\tau=500$, $\gamma=1$, $k=5$, $T=1$.} 
\label{Svsgama_a}
%\vspace{-3mm}
\end{figure}

\begin{figure}[!tbhp]
\centering
\includegraphics[height=6cm,width=7.5cm,angle=0]{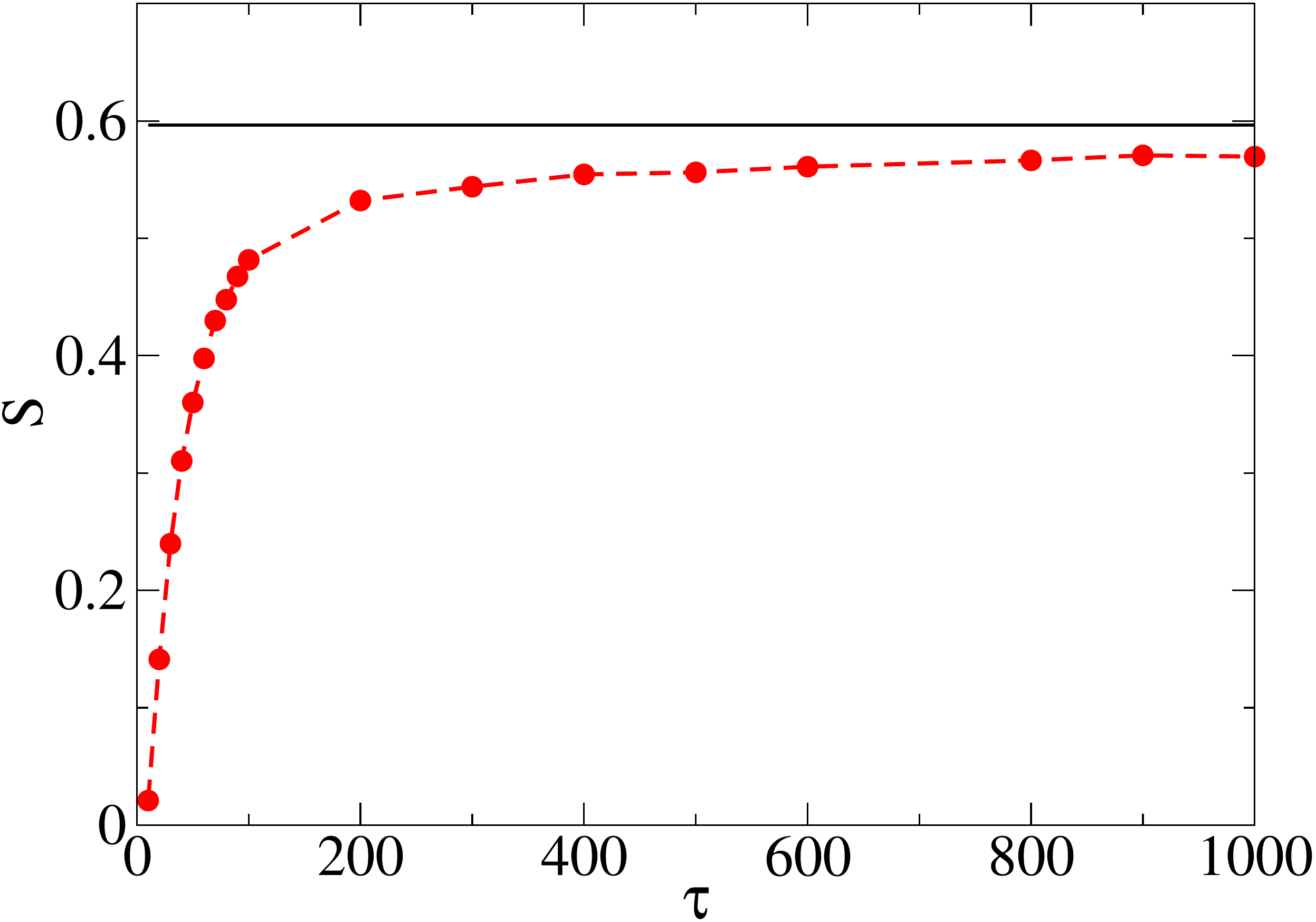}
\caption{Plot of change of entropy along a cycle versus $\tau$. Solid line is the limiting value for long cycle time. Here $\gamma=1$, $\gamma_a=50$, $k=5$, $T=1$.} 
\label{Svst}
%\vspace{-3mm}
\end{figure}

Also, the total average work is same as the total average heat exchanged between the system and the bath, except for a negative sign, i.e., 
$\langle Q\rangle = -\langle W\rangle$. This implies the change of total entropy in a cycle (entropy production) in this limit as, $S=\frac{\langle Q\rangle}{T}=\frac{\langle Q_1-Q_3\rangle}{T}=-\frac{\langle W\rangle}{T}$. Substituting the values of $\langle Q_1\rangle $ and total work $\langle W\rangle $ obtained above. We get:
\beqa
S= -\frac{k_B}{2~(\gamma + \gamma_a)}\ \left[\left(\frac{\gamma - \gamma_a}{2}\ \right) - \gamma_a \ln(2) \right].
\label{entropy}
\eeqa
 Hence, in the limit $\gamma_a >> \gamma $ we get:
\beqa
S= \frac{k_B}{4} (1+2\ln(2))\simeq 0.6k_B
\eeqa 
In  Fig. \ref{Svsgama_a} we have plotted $S$ with $\gamma_a$ for long cycle time obtained analytically as well as from simulation. Interestingly, even in long cycle time limit, unlike Carnot cycle, finite entropy is produced. This indicates non-quasistatic nature of the system even in long $\tau$ limit. In Fig. {\ref{Svst}} we have plotted $S$ with $\tau$ from simulation together with analytic value of $S$. It shows how $S$ approaches towards its limiting value with increasing cycle time.  

\begin{figure}[!tbhp]
\centering
\includegraphics[height=6cm,width=7.5cm,angle=0]{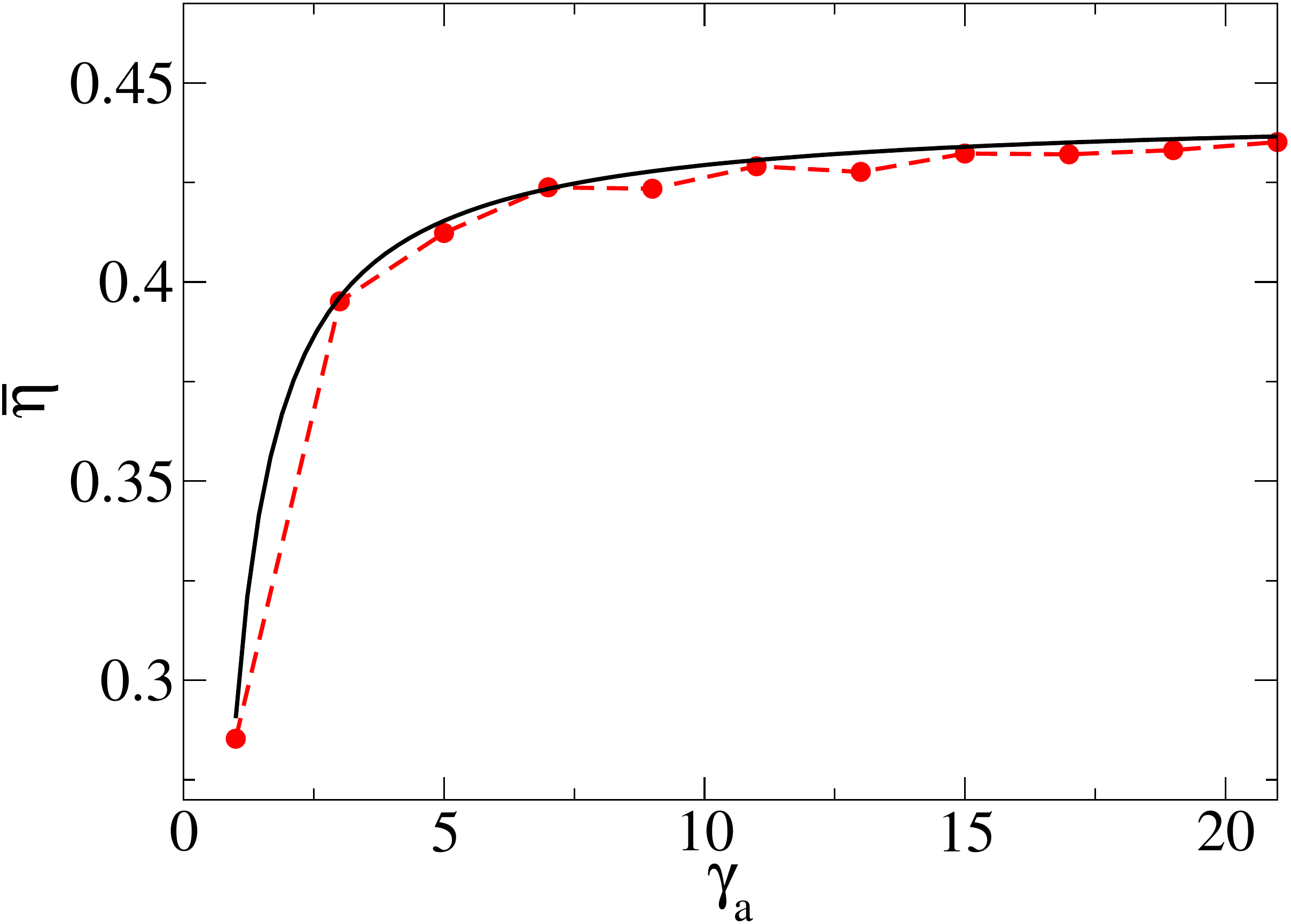}
\caption{Plot of efficiency $\bar\eta$ in long cycle time limit as a function of $\gamma_a$. As $\gamma_a$ increases we recover the value $0.44$ as discussed in the text.
Solid line represent the analytical result namely Eq. $(\ref{etabar})$. Here $\tau=500$, $\gamma=1$, 
$k=5$, and $T=1$.} 
\label{etavsg}
%\vspace{-3mm}
\end{figure}

We can also calculate average thermodynamic efficiency of the machine $\bar\eta$, as defined in Eq. (\ref{etadef}), using average heat in-take of the system along the path of isothermal expansion and the average work extracted from the system along one cycle. 
Now, using expression for $\langle Q_1\rangle$ and $\langle W \rangle$, one can obtain $\bar\eta$ in large $\tau$ limit as:
\beqa
\bar\eta=\frac{\left|\left(\frac{\gamma-\gamma_a}{2}\right) -\gamma_a \ln(2)\right|}{(2\gamma_a-\gamma)+(\gamma+\gamma_a)\ln(2)} \ .
\label{etabar}
\eeqa
For $\gamma_a >> \gamma$ we have, $\bar\eta \simeq \frac{\ln(2)+\frac{1}{2}}{\ln(2)+2}\simeq 0.44$. Note that for $\gamma_a\rightarrow 0$ the device is not acting as an engine. This is plotted in Fig. \ref{etavsg} with $\gamma_a$ and compared to simulations for long cycle time. We note here the fact that the average total entropy production $S$ remains always positive with different cycle times which supports the second law.

For our model, as discussed before, an effective $T_{eff}$ can be defined to restore FDR. If one calculate the efficiency in terms of $T_{eff}$  defined during the second isotherm as, $T_{eff} = T\frac{\gamma}{\gamma+\gamma_a}\ $, which represents the cold temperature in usual Carnot cycle, the corresponding Carnot efficiency is $\eta_c = 1-\frac{T_{eff}}{T_h}\ = 1- \frac{\gamma}{\gamma+\gamma_a}\ $. However, this efficiency $\rightarrow 1$ for $\gamma_a >> \gamma$. Thus it is almost double than $\bar\eta$ we obtained above.    

\subsection{Statistics Of Stochastic Thermodynamic Quantities : Distributions and Large Deviation Functions}

\begin{figure}[htp]
\centering
\includegraphics[width=0.85\columnwidth]{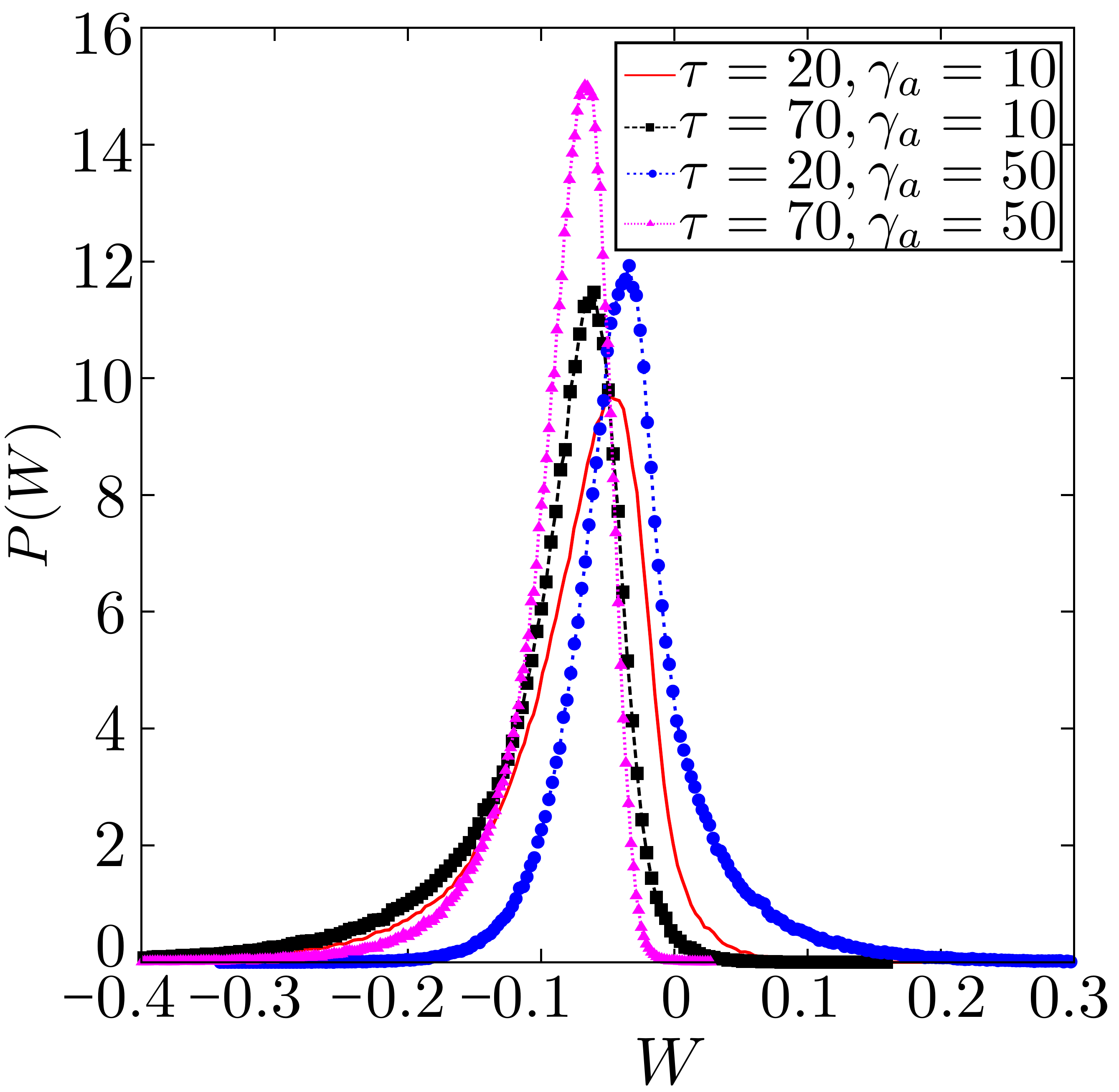}
\caption{ Plot of work probability densities for various combinations of $\tau$ and $\gamma_a$.}
\label{pw}
\vspace{-3mm}
\end{figure}

Though on an average, work is extracted from the bath, fluctuations dominate and $W$ follows a broad probability density function as seen in Fig. \ref{pw}, where work probability densities for different combinations of $\tau$ and $\gamma_a$ are calculated by simulating the system.

\begin{figure}[!tbhp]
\centering
\hspace{-0.5cm}
\includegraphics[height=5.5cm, width=8cm, angle=0]{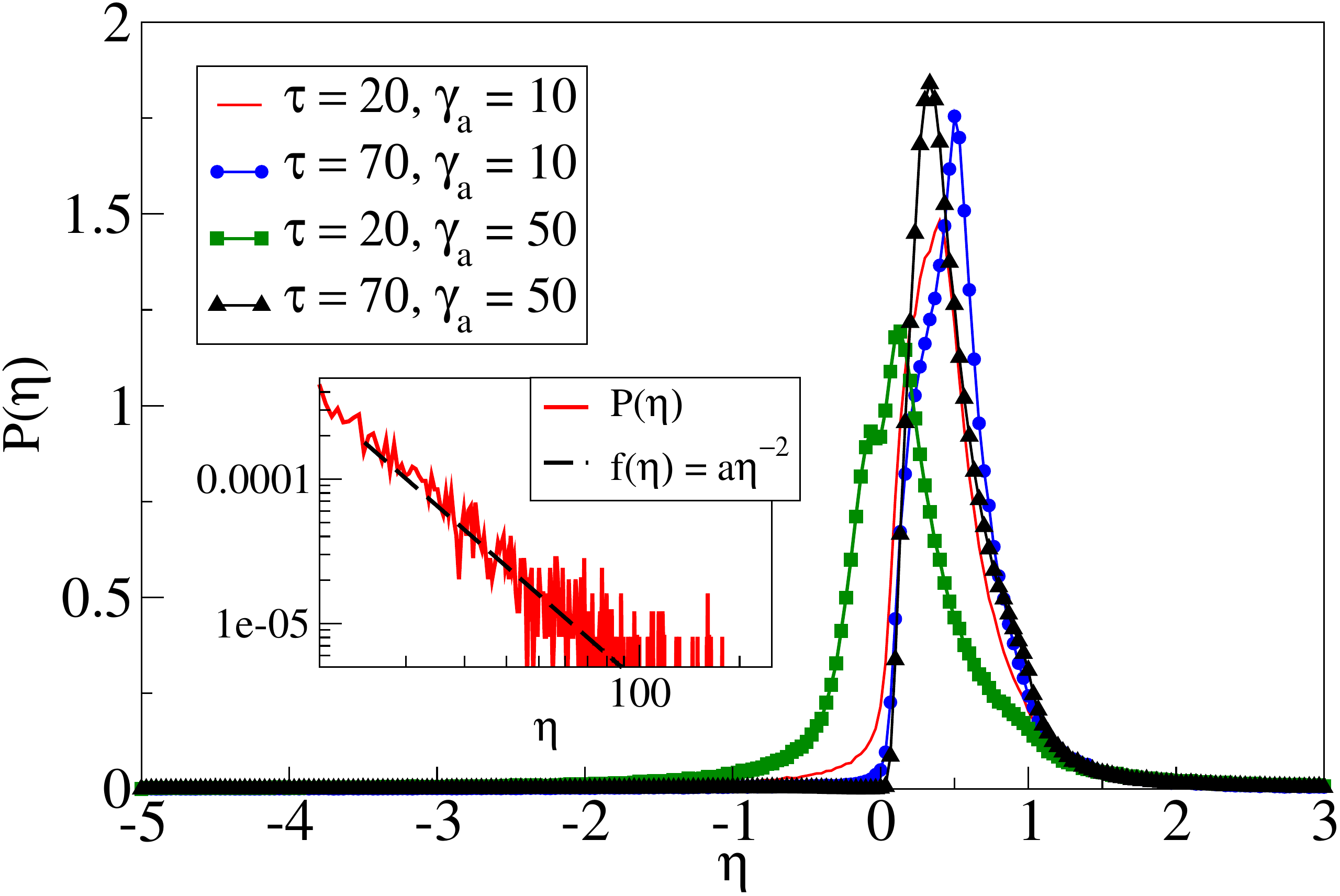}
\caption{Plot of probability density of efficiency $P(\eta)$ vs $\eta$ for different values of $\gamma_{a}$ and $\tau$. Distribution has power law tails with exponent close to $2$ $($inset$)$.} 
\label{peta}
%\vspace{-3mm}
\end{figure}

In Fig. \ref{peta}, efficiency distribution $P(\eta)$ is shown for different combinations of $\gamma_{a}$ and $\tau$. In the inset we have shown that the tail of $P(\eta)$ goes as $\eta^{-\alpha}$ with $\alpha\simeq 2$. The stochastic efficiency is unbounded and distribution is very broad. It shows power law tail with exponent around $2$. Fluctuations in $\eta$ are large \cite{Arun14,ArunPhysicaA,ArunIJMPB}. In fact, relative variance of the stochastic efficiency is much larger than mean value. This implies that the average quantity is not a good physical variable here. In such situations one has to study the full probability distribution of $\eta$. However, one can quantify the most probable value of efficiency. A suitable characterization, where this value is enhanced is the large deviation function, which we will calculate next.

In the limit of large observation time $($$\tau_{obs}$$)$ the distribution of stochastic efficiency is characterized by its large deviation function (LDF) $J_{\tau}(\eta')$, defined as \cite{Martinez16}: 

\begin{equation}
 P^{(\tau_{obs})}_{\tau}(\eta')\simeq \exp\left[-\tau_{obs}J_{\tau}(\eta')\right],\hspace{0.5 cm}\tau_{obs}\rightarrow\infty,
 \label{pdfldf}
\end{equation}
here subscript $\tau$ indicates the period of one engine cycle. The cycle time $\tau$ and the 
observation time $\tau_{obs}$ are related by $\tau_{obs}=M\times\tau$, where $M$ is the number of 
cycles over which efficiency has been calculated. Here $P^{(\tau_{obs})}_{\tau}(\eta')$ is the 
probability distribution function of the efficiency $\eta'$, obtained by summing the works and the heats for all the $M$ cycles and then taking the ratio of these sums:
\begin{equation}
 \eta'=\frac{\sum_{i=1}^M W_i}{\sum_{i=1}^M Q_i}.
\end{equation}
From Eq. $(\ref{pdfldf}$ one can estimate the LDF corresponding to the efficiency distribution as:
\begin{equation}
 J_{\tau}(\eta')\simeq -\lim_{ \tau_{obs}\rightarrow\infty}\frac{1}{\tau_{obs}}\ln P^{(\tau_{obs})}_{\tau}(\eta').
\end{equation}

\begin{figure}[!tbhp]
\centering
\includegraphics[height=5.5cm, width=8cm, angle=0]{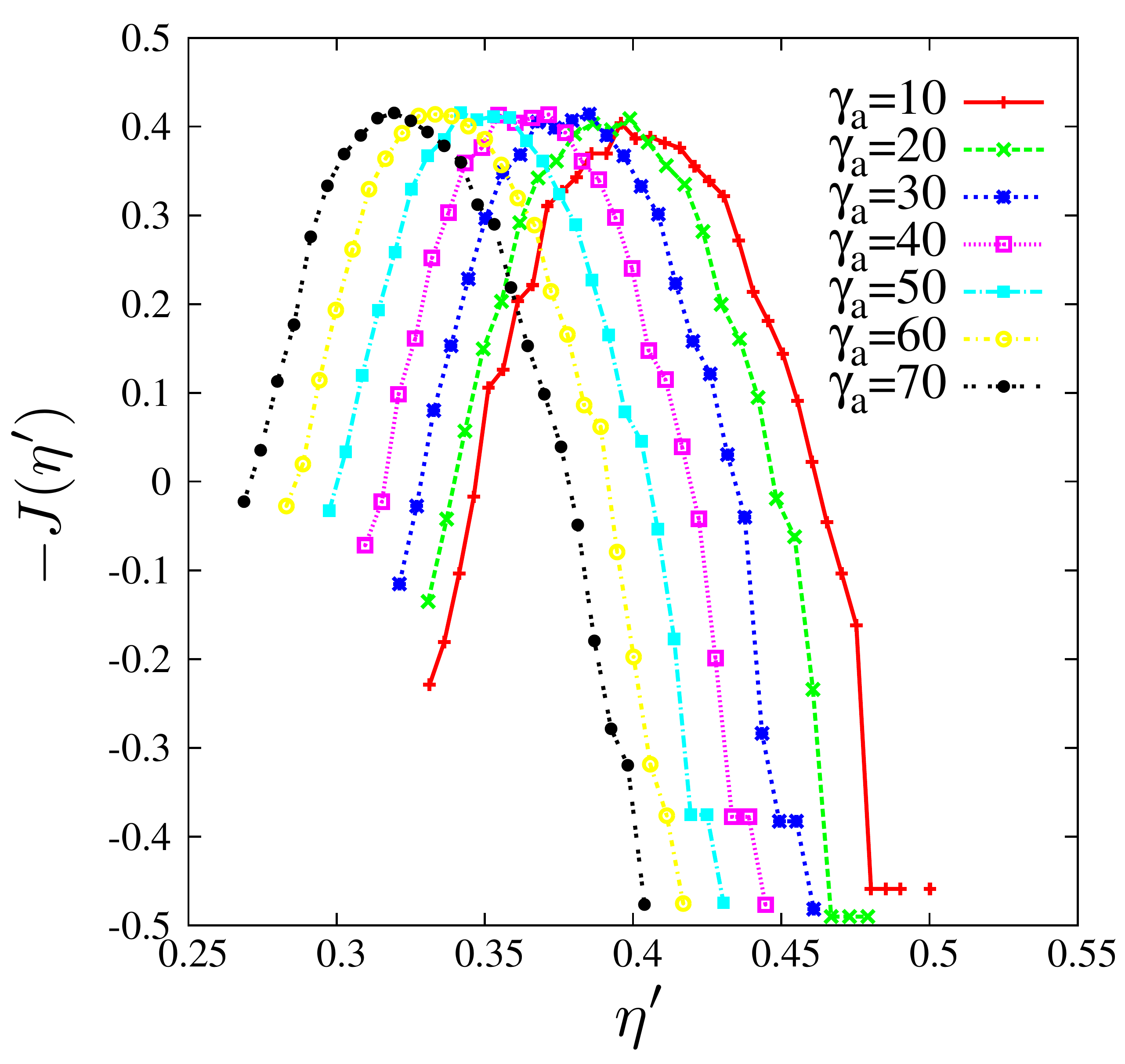}
\caption{Plot of negative of large deviation function for different values of $\gamma_{a}$ with $\tau=70$. Here $\tau_{obs}=100 \times \tau$.}
\label{ldf}
%\vspace{-3mm}
\end{figure}

In Fig. \ref{ldf}, we have plotted $-J_{\tau}(\eta')$ as a function of efficiency $\eta'$ for different values of $\gamma_a$. It shows a maximum for every value of $\gamma_a$. This implies that in the limit of large observation time the efficiency value corresponding to the maximum is the most probable value in the efficiency statistics. Interestingly, the most probable efficiency decreases as one approaches to higher values of $\gamma_a$.


\section{ Conclusion} 

We have considered a single Brownian particle kept in a time-dependent harmonic trap as discussed in reference \cite{Arun14}. Though in the model discussed here, instead of two heat baths we have a single bath at temperature $T$. However during the compression step, the frictional drag $\gamma v$ provided by the bath to the particle has been increased to $(\gamma + \gamma_{a})v$ with $(\gamma,\gamma_{a}) > 0$. Therefore, the effective friction coefficient during the compression is large compared to that of in the expansion process. Though the temperature of the bath is constant through out the dynamics, due to the effective, large friction, the heat loss during compression is more than the expansion process. So, to the particle, the effective temperature $T_{{eff}}$ of the surroundings appears to be smaller than $T$. Hence we can extract work from single heat bath without violating the second law of thermodynamics. 

This technology is extremely important which allows extracting work from a single bath by an open-loop protocol. Similar technology where, instead friction, thermal noise has been enhanced by adding electrical noise to imitate the bath of higher temperature of a Carnot-type micro-heat engine, has recently been realized experimentally \cite{EdgarAPP,Martinez16}. In this case, additional electrical noise breaks FDR along the hot isotherm. 

We have mentioned here that the fluctuations from out-of-equilibrium bath (e.g. bacterial bath) can be considered as thermal fluctuations with an {\it effective temperature} different from the existing bath temperature. In the model concerned here, $T_{{eff}}$ that restores FDR in the compression step of the cycle is $\frac{T\gamma}{\gamma+\gamma_a}$. We should also mention here that the consideration of effective temperature is debatable in many out-of-equilibrium systems \cite{Frank2001,Palacci2010,Ngov2011,Prost2005} and to address this issue properly one might need to explore the thermostatistics of a passive tagged particle immersed in an active bath in detail which is beyond the scope of this paper.

The protocol can be experimentally realized by the usage of photoactive, self-propelled, micro-entities as bath-particles ( e.g. bacterial bath, bath of active colloids etc.) where one may tune the drag force on the Brownian particle by tuning the activity (i.e. self-propulsion), keeping the thermal fluctuations of the bath unaltered. Importantly, as we have discussed earlier in this article, this will eventually provide a novel control on the most probable efficiency of the micro machines. 

We also note that the power-law exponent $\alpha\simeq 2$ obtained from the tail of $P(\eta)$ here, is also obtained in various other micro machines as, $(i)$ in \cite{VerleyPRL, Arun14, ArunPhysicaA, ArunIJMPB} $(ii)$ in case of a classical spin-$1/2$ system coupled to two heat baths simultaneously \cite{basu} and $(iii)$ in case of a micro-heat engine with a Brownian particle driven by micro adiabatic protocol \cite{Edgar16,saha}.  The performance of the engine is dominated by fluctuations and hence it is not a reliable engine. It will be interesting to extract work from non-equilibrium bath, as in the present case, however with an optimal protocol. This should enhance the performance characteristics of such engines.

\section{Author Contribution}

AS and RM have contributed equally to this work.

\section{Acknowledgments}
AS thanks University Grants Commission Faculty Recharge Program (UGCFRP), India and 
RM thanks Department of Science and Technology (DST), India for financial support. 
AMJ also thanks DST, India for J. C. Bose National Fellowship. AS thanks Edgar Roldan for 
initial discussions on micro heat engines.

%---------------------------------------------------------------------------------%

%---------------------------------------------------------------------------------%
\end{document}